\documentclass[twocolumn,aps,prc,showpacs,amssymb,floatfix,10pt]
{revtex4}
\usepackage{exscale,times} 
\usepackage{bm}                      
\usepackage[dvips]{graphicx}         
\usepackage{dcolumn}                 

\begin{document}
\def\bea{\begin{eqnarray}}
\def\eea{\end{eqnarray}}
\def\be{\begin{equation}}
\def\ee{\end{equation}}
\def\rra{\right\rangle}
\def\lla{\left\langle}
\def\rv{\bm{r}}
\def\la{\Lambda}
\def\sgm{\Sigma^-}
\def\eps{\epsilon}
\def\ms{M_\odot}
\def\bc{B=100\;\rm MeV\!/fm^3}
\def\beff{B_{\rm eff}(\rho_B)}
\def\sc{\sigma_{\rm crit} \approx 70\;\rm MeV\!/fm^2} 
\def\qc{\rho_{\rm ch}}
\def\fv{f_V}


\title{Hadron-quark mixed phase in hyperon stars}

\author{
Toshiki Maruyama,$^1$ 
Satoshi Chiba,$^1$ 
Hans-Josef Schulze,$^2$ and
Toshitaka Tatsumi$^3$
}

\affiliation{
$^1$ Advanced Science Research Center, Japan Atomic Energy Agency,
     Tokai, Ibaraki 319-1195, Japan \\
$^2$ INFN Sezione di Catania, 
     Via Santa Sofia 64, I-95123 Catania, Italy \\
$^3$ Department of Physics, Kyoto University, Kyoto 606-8502, Japan
}


\begin{abstract}
We analyze the different possibilities for the hadron-quark phase transition
occurring in beta-stable matter including hyperons in neutron stars.
We use a Brueckner-Hartree-Fock approach including hyperons for the hadronic
equation of state and a generalized MIT bag model for the quark part.
We then point out in detail the differences between Maxwell and Gibbs phase
transition constructions
including the effects of surface tension and electromagnetic screening.
We find only a small influence on the maximum neutron star mass,
whereas the radius of the star and in particular its internal structure
are more affected.
\end{abstract}


\maketitle

\section{Introduction}

The theoretical understanding of neutron star (NS) structure requires 
the knowledge of the equation of state (EOS) of highly compressed cold
baryonic matter, up to densities of about ten times normal nuclear density,
$\rho_0 \approx 0.17\;\rm fm^{-3}$ \cite{ns}.
In such an extreme environment, the appearance of ``exotic'' components
of matter, such as hyperons, meson condensates, and quark matter (QM),
is expected \cite{nshyp}.

It is in fact well known that hyperons appear at around \hbox{2--3$\,\rho_0$}
in beta-stable nuclear matter and lead to a strong softening of the EOS
with a consequent substantial reduction
of the maximum NS mass \cite{gle,glenhyp}.
The theoretical maximum mass of hyperonic NS can even result below current 
observational values of about 1.5 solar masses \cite{nsmass},
as in the case of the microscopic Brueckner-Hartree-Fock (BHF) approach
for the hyperonic EOS \cite{hypns,hypns2}, 
which we employ in this work for the hadronic phase.
This would imply the presence of nonhadronic ``quark'' matter in the 
interior of massive NS \cite{mit,njl,cdm,njlc},
heavier than the maximum mass hyperon stars.

However, the appearance of quark matter poses the problem of
an accurate theoretical description of the quark phase,
which is so far an open question,
and furthermore of the
details of the phase transition between hadronic and quark matter.
The purpose of this article is the study of the latter problem,
combining the BHF EOS of hyperonic matter with a generalized
phenomenological MIT bag model for the quark phase.
In our case 
the EOS of hyperonic matter is so soft that the deconfinement transition 
occurs at rather low densities, where hyperon contamination is not so large. 
Assuming the quark deconfinement transition to be of first order, 
it causes a thermodynamical instability and the mixed phase (MP) appears
around the critical density. 
In the usual Maxwell construction (MC),
the MP is composed of two charge-neutral hadron and quark phases
and uniform density distributions are assumed in each phase.

The properties of the MP are very different,
if there are more than one independent chemical potentials, as in our case.
Glendenning has pointed out that in this case the MC
is not the correct procedure to obtain a thermodynamically well-defined EOS
with the MP, but that one must properly satisfy more fundamental 
requirements by means of the Gibbs conditions \cite{glen}. 
Since then many works have appeared regarding 
nuclear pasta structures in low-density nuclear matter 
\cite{mixmaru,mixmarurev},
kaon condensation at several times $\rho_0$ \cite{mixmaru,mixmarurev,chr,nor},
and the hadron-quark deconfinement transition
\cite{glen,mixmarurev,mix,vos,mixtat}.
When the Gibbs conditions are applied to the quark deconfinement transition,
the MP is composed of individually charged hadron and quark phases,
and baryon number density as well as charge density
are nonuniform in each phase,
arranged in different geometrical structures.

However, electromagnetic and surface contributions
to the energy of the MP are only approximately 
treated in the usual bulk calculations \cite{mix}, 
but could have an important effect \cite{mixmarurev,vos,mixtat,nor}.
The quantitative analysis of these corrections
(sometimes called {\em finite-size effects}) for the quark
deconfinement transition in hyperonic matter is one purpose of this article.
We elaborately figure out the roles of the finite-size effects in the
mixed phase.
We shall see that these effects change remarkably the properties
of the mixed phase; 
e.g., the geometrical structures are destabilized by the
charge screening effect for the Coulomb interaction in the extreme case, 
and the EOS given by the MC is effectively recovered. 
Regarding hyperon mixing we shall see that the appearance of the 
mixed phase completely suppresses the appearance of hyperons. 

Some major results of our work were already presented in another 
paper \cite{let}, while here we provide more details 
and furthermore study the influence of the mixed phase 
on the global NS observables like mass-radius relation and in particular
the maximum mass. 
We shall see that the global properties of compact stars are little changed, 
but the structure and property of the internal core are very different, 
compared to the MC. 
Such difference may affect the elementary processes like neutrino transport 
or neutrino emission in the core. 

In the following, 
we give in Sec.~\ref{s:bhf} a concise summary of the BHF approach for
hyperonic matter that is used, 
whereas in Sec.~\ref{s:mit} the modified MIT bag model for the quark 
phase is introduced.
Sec.~\ref{s:mix} discusses the details of 
the properties of the mixed phase and figures out the peculiar role of
the finite-size effects.
In Sec.~\ref{s:ns} the EOS including the quark deconfinement transition is
applied to the structure of hybrid stars.
Summary and concluding remarks are given in Sec.~\ref{s:end}.

\section{BHF approach of hyperonic matter}
\label{s:bhf}

Our theoretical framework for the hadronic phase of matter
is the nonrelativistic Brueckner-Hartree-Fock approach \cite{bhf}
based on microscopic 
nucleon-nucleon (NN), nucleon-hyperon (NY), and hyperon-hyperon (YY) 
potentials that are fitted to scattering phase shifts, where possible.
Nucleonic three-body forces (TBF) are included in order to (slightly) shift
the saturation point of purely nucleonic matter to the empirical value.

It has been demonstrated that the theoretical basis of the BHF method, 
the hole-line expansion, is well founded:
the nuclear EOS can be calculated with good accuracy in the BHF two hole-line 
approximation with the continuous choice for the single-particle potential, 
since the results in this scheme are quite close to the full
convergent calculations which include also the three hole-line
contributions \cite{thl}.
Due to these facts, combined with the absence of adjustable parameters, 
the BHF model is a reliable and well-controlled theoretical approach 
for the study of dense baryonic matter.

In the following we give a short review of the BHF approach including hyperons.
Detailed accounts can be found in Refs.~\cite{hypns,hypmat,vi00}.
The basic input quantities in the Bethe-Goldstone equation
are the NN, NY, and YY potentials. 
In this work we use the Argonne $V_{18}$ NN potential \cite{v18} supplemented by
the Urbana UIX nucleonic TBF of Ref.~\cite{uix}
and the Nijmegen soft-core NSC89 NY potentials \cite{nsc89}
that are well adapted to the existing experimental NY scattering data
and also compatible with $\la$ hypernuclear levels \cite{yamamoto,vprs01}.
Unfortunately, the NSC89 potentials contain no YY components,
because up to date no YY scattering data exist. 
Nevertheless the importance of YY potentials should be minor 
as long as the hyperonic partial densities remain limited.
Recently also calculations using the NSC97 NY and YY potentials \cite{nsc97}
were completed, which yield very similar maximum NS masses in spite 
of quite different internal compositions \cite{hypns2}.

With these potentials, the various $G$ matrices are evaluated 
by solving numerically the Bethe-Goldstone equation, which can be written in 
operator form as 
\be
 G_{ab}[W] = V_{ab} + \sum_c \sum_{p,p'} V_{ac} \big|pp'\big\rangle 
 {Q_c \over W - E_c +i\eps} 
  \big\langle pp'\big| G_{cb}[W] \:, 
\label{e:g}
\ee
where the indices $a,b,c$ indicate pairs of baryons
and the Pauli operator $Q$ and energy $E$ 
determine the propagation of intermediate baryon pairs.
The pair energy in a given channel $c=(ij); i,j=n,p,\la,\sgm$ is
\be
 E_{(ij)} = T_{i}(k_{i}) + T_{j}(k_{j})
 + U_{i}(k_{i}) + U_{j}(k_{j})
\label{e:e}
\ee
with 
$T_i(k) = m_i + {k^2\!/2m_i}$,
where the various single-particle potentials are given by
\be
 U_i(k) = 
 \sum_{j=n,p,\la,\sgm} U_i^{(j)}(k)
\label{e:un}
\ee
and are determined self-consistently from the $G$ matrices,
\be
  U_i^{(j)}(k) = 
  \!\!\! \sum_{k'<k_F^{(j)}} \!\!\!
  {\rm Re} \big\langle k k' \big| G_{(ij)(ij)}\big[E_{(ij)}(k,k')\big] 
  \big| k k' \big\rangle  \:.
\label{e:uy}
\ee
The coupled equations (\ref{e:g}) to (\ref{e:uy}) define the BHF scheme 
with the continuous choice of the single-particle energies.  
In contrast to the standard purely nucleonic calculation,
the additional coupled-channel structure renders the calculations 
quite time-consuming.

Once the different single-particle potentials are known,
the total nonrelativistic hadronic energy density, $\eps_H$,  
can be evaluated:
\bea
 \eps_H \!&=& \!\!\!\!
 \sum_{i=n,p,\la,\sgm} 
 \sum_{k<k_F^{(i)}}
 \left[ T_i(k) + {1\over2} U_i(k) \right] \:,
\label{e:eps}
\eea
and $\eps_H$ is thus represented as a function of particle number densities
$\rho_i (i=n,p,\la,\sgm)$ for a given baryon number density $\rho_B$. 
Knowing the hadronic energy density, 
and adding the contributions of the noninteracting leptons ($l=e,\mu$),
$\eps=\eps_H+\eps_L$,
the various chemical potentials 
$\mu_i = \partial\eps / \partial\rho_i$
(of the species $i=n,p,\la,\sgm,e,\mu$)
can be computed straightforwardly
and the equations for beta-equilibrium,
\be
 \mu_i = B_i \mu_n - Q_i \mu_e \:,
\label{e:mu}
\ee
($B_i$ and $Q_i$ denoting baryon number and electric charge of species $i$),
baryon number density and charge neutrality,
\bea
 \sum_i B_i \rho_i &=& \rho_B \:,
\\
 \sum_i Q_i \rho_i &=& 0 \:,
\eea
allow one to determine the equilibrium composition $\rho_i(\rho_B)$
at a given baryon number density $\rho_B$ 
and finally the hadronic EOS,
\be
 P_H(\rho_B) = \rho_B^2 {d\over d\rho_B} 
 {\eps(\rho_i(\rho_B))\over \rho_B}
 = \rho_B {d\eps \over d\rho_B} - \eps 
 \:.
\ee

\section{Quark matter EOS}
\label{s:mit}

Unfortunately, the current theoretical description of quark matter 
is burdened with large uncertainties, seriously limiting the 
predictive power of any theoretical approach at high baryonic density. 
For the time being we can therefore only resort
to phenomenological models for the quark matter EOS 
and try to constrain them as well as possible 
by the few experimental information available on high-density baryonic matter.

One important condition is due to the fact 
that certainly in symmetric nuclear matter no phase transition is observed  
below $\approx 3\rho_0$.
We will in the following use an extended MIT bag model \cite{chodos}
(requiring a density-dependent bag ``constant'')
that is compatible with this condition.

\subsection{The MIT bag model}

We first review briefly the description of the bulk properties of 
uniform quark matter,
deconfined from the beta-stable hadronic matter mentioned in the
previous section, by using the MIT bag model \cite{chodos}.
The energy density of the $f=u, d, s$ quark gas 
can be expressed as a sum of the kinetic term
and the leading-order one-gluon-exchange term \cite{quark,jaf}
for the interaction energy
proportional to the QCD fine structure constant $\alpha_s$, 
\bea
 \epsilon_Q &=& B + \sum_f \epsilon_f \:,
\\
 \epsilon_f(\rho_f) &=& {3m_f^4 \over 8\pi^2} \bigg[ 
 { x_f\left(2x_f^2+1\right)\sqrt{1 + x_f^2}} - {\rm arsinh}\,x_f \bigg] 
\nonumber \\
 && - \alpha_s{m_f^4\over \pi^3} \bigg[ 
 x_f^4 - {3\over2}\Big( x_f \sqrt{1 + x_f^2} - {\rm arsinh}\,x_f \Big)^2 
 \bigg] \:,
\nonumber \\&&
\eea
where $m_f$ is the $f$ current quark mass,
$x_f = k_F^{(f)}\!/m_f$,
the number density of $f$ quarks is $\rho_f = {k_F^{(f)}}^3\!\!/\pi^2$,
and $B$ is the energy density difference between 
the perturbative vacuum and the true vacuum, i.e., the bag ``constant.''
We use massless $u$ and $d$ quarks and $m_s= 150$ MeV.

It has been found \cite{mit,alford} that within the MIT bag model
(without color superconductivity) with a density-independent bag
constant $B$, the maximum mass of a NS cannot exceed a value of
about 1.6 solar masses. 
Indeed, the maximum mass increases as the
value of $B$ decreases, but too small values of $B$ are incompatible
with a hadron-quark transition density $\rho_B >$ 2--3 $\rho_0$ in 
nearly symmetric nuclear matter, 
as demanded by heavy-ion collision phenomenology. 
Values of $B\gtrsim 150\;\rm MeV\!/fm^{3}$ 
can also be excluded within our model, 
since we do not obtain any more a phase transition in beta-stable matter 
in combination with our hadronic EOS \cite{mit}.

In order to overcome these restrictions of the model, 
one can introduce empirically a density-dependent bag parameter $\beff$, 
which has not any more the simple interpretation as the energy difference
between the perturbative and the true vacua; 
instead some density dependence originating from the non-perturbative
interaction energy $\Delta\epsilon_{\rm int}(\rho_B)$ 
is considered to be included in the effective bag constant, i.e., 
$\beff = B + \Delta\epsilon_{\rm int}(\rho_B)$ \cite{cdm}. 
This allows one to lower the value of $B$ at large density, 
providing a stiffer QM EOS and increasing the value of the maximum NS mass, 
while at the same time still fulfilling the condition of no phase 
transition below $\rho_B \approx 3 \rho_0$ in symmetric matter.
In the following we present results based on the MIT model using both 
a constant value of the bag parameter, $\bc$, and a
Gaussian parametrization for the density dependence,
\be
 \beff = B + (B_0 - B)
 \exp\left[-\beta\Big(\frac{\rho_B}{\rho_0}\Big)^2 \right]
\label{eq:param} 
\ee 
with $B= 50\;\rm MeV\!/fm^{3}$, 
$B_0 = 400\;\rm MeV\!/fm^{3}$, and $\beta=0.17$.
For a more extensive discussion of this topic, the reader is referred to 
Refs.~\cite{mit,njl,cdm}.

The introduction of a density-dependent bag parameter 
has to be taken into account properly for the 
computation of various thermodynamical quantities; 
in particular the quark chemical potentials
are modified as
\bea
 \mu_f &\rightarrow& \mu_f + {1\over3}{d\beff \over d\rho_B} \:.
\label{e:murho}
\eea
Nevertheless, due to a cancellation of the second term in (\ref{e:murho}), 
occurring in relations (\ref{e:chem}) for the beta-equilibrium, 
the composition of QM at a given total baryon number density remains unaffected
by this term (and is in fact independent of $B$).
At this stage of investigation, we disregard possible dependencies 
of the bag parameter on the individual quark densities. 

In the beta-stable pure quark phase,
the individual quark chemical potentials
are fixed by Eq.~(\ref{e:mu}) with $B_f=1/3$, 
which implies
\be
 \mu_d = \mu_s = \mu_u + \mu_l \:.
\label{e:chem} 
\ee
The charge neutrality condition and the total
baryon number conservation read
\bea
 0 &=& \frac{2}{3}\rho_u - \frac{1}{3}\rho_d - \frac{1}{3}\rho_s - \rho_l \:,
\\
 \rho_B &=& \frac{1}{3}\left(\rho_u + \rho_d + \rho_s\right) \:.
\label{e:baryon} 
\eea
These equations determine the composition $\rho_f(\rho_B)$
and the pressure of the QM phase,
\be
 P_Q(\rho_B) = \rho_B^2 {d(\eps_Q/\rho_B) \over d\rho_B}\:.
\ee

The modified bag model is clearly an oversimplified model of QM,
but in this article we focus mainly on the differences
between the different hadron-quark phase transition constructions 
in NS matter introduced in the following.

\section{Hadron-quark phase transition}
\label{s:mix}

\subsection{Gibbs conditions and the mixed phase}

\begin{figure}
\includegraphics[width=0.48\textwidth]{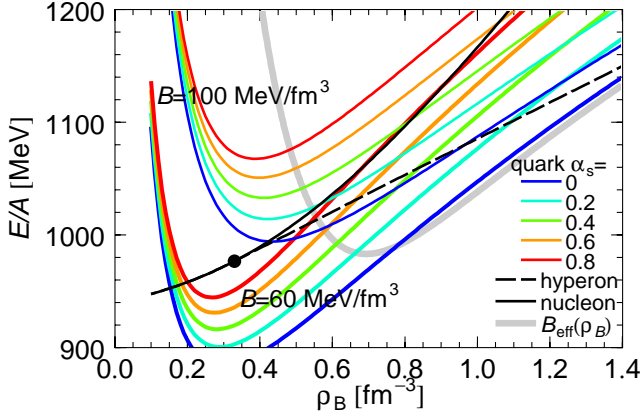}
\caption{
(Color online)
EOS of beta-stable hadronic matter (black curves)
and quark matter (colored curves)
with constant $B=60\;\rm MeV\!/fm^3$ (lower curves)
and $B=100\;\rm MeV\!/fm^3$ (upper curves) 
for several values of $\alpha_s$.
The gray curve shows the result for the $\beff$ model with $\alpha_s=0$.
Hyperons appear at the dotted point in hadronic matter.}
\label{figParam}
\end{figure}

Figure~\ref{figParam} compares the hadronic BHF EOS 
and the quark matter EOS with 
different values of the parameters $B$ and $\alpha_s$
for beta-stable and charge-neutral matter.
One can see that the quark EOS approaches that of a relativistic
free gas ($E/A \sim \rho_B^{1/3}$) with increasing density, 
while the hyperonic EOS is always soft.
Consequently the quark deconfinement transition 
cannot occur at too high densities.
If we demand the quark and the hyperonic EOS to cross,
$\alpha_s$ should be small and $B$ slightly large, 
which gives a relatively low critical density.
Thus the appearance of hyperons is effectively suppressed 
due to a quark deconfinement transition.
In this article we choose $\alpha_s=0$ and 
compare results with $\bc$ and $\beff$.

Since we shall see that two independent chemical potentials, 
charge chemical potential and baryon number chemical potential, 
are needed to specify the thermodynamical properties of the MP, 
we must properly take into account the Gibbs conditions \cite{mix},
which require the pressure balance and the equality of the chemical potentials
between the two phases besides the thermal equilibrium.
We employ a Wigner-Seitz approximation in which the whole space 
is divided into equivalent cells with a given geometrical symmetry, 
specified by the dimensionality 
$d=3$ (sphere), $d=2$ (rod), or $d=1$ (slab).
In each cell with volume $V_W$ the quark and hadron phases are separated: 
a lump portion made of the quark phase with volume $V_Q$ is embedded
in the hadronic phase with volume $V_H$ or vice versa.
A sharp boundary is assumed between the two phases and the surface energy
is taken into account in terms of a surface-tension parameter $\sigma$.
The surface tension of the hadron-quark interface is poorly known, 
but some theoretical estimates based on the MIT bag model 
for strangelets \cite{jaf} and
lattice gauge simulations at finite temperature \cite{latt} suggest
a range of $\sigma \approx 10-100\;\rm MeV\!/fm^2$.
We show results using $\sigma=40\;\rm MeV\!/fm^2$ 
in this article, and discuss the effects of its variation.

We use the Thomas-Fermi approximation for the density profiles of
hadrons and quarks.
The energy for each cell is then given as
\be
 E = \int_{V_H}\!\!\! d^3\rv\, \eps_H[\rho_i(\rv)] 
 + \int_{V_Q}\!\!\! d^3\rv\, \eps_Q[\rho_f(\rv)] + E_e + E_C + \sigma S
\ee
with $i=n,p,\la,\sgm$, $f=u,d,s$, 
and $S$ being the hadron-quark interface area. 
$E_e$ is the electron gas kinetic energy,
\be
 E_e = \int_{V_W}\!\!\! d^3\rv\, \eps_e[\rho_e(\rv)] \:,
\ee
approximated by 
$\eps_e[\rho_e] \simeq (3\pi^2\rho_e)^{4/3}\!/4\pi^2$.
For simplicity muon is not included in this article.
$E_C$ is the Coulomb interaction energy,
\be
 E_C = \frac{e^2}{2} \int_{V_W}\!\!\! d^3\rv d^3\rv'\,
 {\qc(\rv)\qc(\rv') \over |\rv-\rv'|} \:,
\ee
where the charge density is given by 
$e\qc(\rv) = \sum_{i=n,p,\la,\sgm,e} Q_i\rho_i(\rv)$ in $V_H$
and 
$e\qc(\rv) = \sum_{f=u,d,s,e} Q_f \rho_f(\rv)$ in $V_Q$  
with $Q_a$ being the particle charge 
($Q=-e < 0$ for the electron).
Accordingly, the Coulomb potential $\phi(\rv)$ is defined as
\be
 \phi(\rv) = -\int_{V_W}\!\!\! d^3\rv'\, 
 {e^2 \qc(\rv') \over \left| \rv - \rv' \right|} + \phi_0 \:,
\label{e:vcoul}
\ee
where $\phi_0$ is an arbitrary constant representing 
the gauge degree of freedom.
We fix it by stipulating the condition 
$\phi(R_W) = 0$, 
as before \cite{vos,mixtat,mixmaru}.
The Poisson equation then reads
\bea
 \Delta \phi (\rv) = 4 \pi e^2 \qc(\rv) \:.
\label{e:poisson}
\eea

The density profiles of the hadrons, 
$\rho_i(\rv),\ i=n,p,\la,\sgm$, 
and quarks, $\rho_f(\rv),\ f=u,d,s$,
are determined by the equations of motion,
\be
 {\delta (E/V_W) \over \delta\rho_a(\rv)} = \mu_a \:,
\label{e:eom}
\ee
where we introduced the chemical potentials $\mu_a$ for the particle species 
$a=n,p,\la,\sgm,u,d,s,e$. 
Note that some additional terms are needed for the 
quark chemical potentials in the case of $\beff$ [see Eq.~(\ref{e:murho})].
We consider chemical equilibrium at the hadron-quark
interface as well as inside each phase, 
so that Eq.~(\ref{e:mu}) implies
\bea
 && \mu_u+\mu_e = \mu_d = \mu_s \:, 
\nonumber\\
 && \mu_p+\mu_e = \mu_n = \mu_\la = \mu_u + 2\mu_d \:, 
\nonumber\\
 && \mu_{\sgm} + \mu_p = 2\mu_n \:.
\label{e:chemeq}
\eea
There are two independent chemical potentials, which are usually chosen as
the charge chemical potential $\mu_Q=\mu_e$ and the baryon number chemical
potential $\mu_B=\mu_n$.
For a given baryon number density 
\be
 \rho_B = {1\over V_W}\! \left[
 \sum_{i=n,p,\la,\sgm} \int_{V_H}\!\!\!\! d^3\rv \rho_i(\rv)
 +\! \sum_{f=u,d,s} \int_{V_Q}\!\!\!\! d^3\rv {\rho_f(\rv)\over3} \right] \:,
\ee
Eqs.~(\ref{e:poisson}--\ref{e:chemeq}), 
together with the global charge neutrality condition,
$\int_{V_W}\!\!{d^3\rv} \qc(\rv)=0$, 
obviously fulfill the requirements by the Gibbs conditions. 
The optimum dimensionality $d$ of the cell or the lump,
the cell size $R_W$, and the lump size $R$ 
[or equivalently the volume fraction $\fv\equiv (R/R_W)^d$],
are searched for to give the minimum energy.

Note that the Poisson equation~(\ref{e:poisson}) is a highly non-linear 
equation with respect to the Coulomb potential through Eq.~(\ref{e:eom}). 
If we linearize it, we obtain the Debye screening length
\be
 {1\over \lambda_D^2} = 4\pi e\sum_{i} 
 Q_i {\partial\langle \qc \rangle \over \partial\mu_i} \:,
\ee
(with
$i=n,p,\la,\sgm,e$ and $i=u,d,s,e$
in the hadron and quark phases, respectively),
which gives a rough measure to estimate how effective is the charge screening. 
We shall see that 
$\lambda_D \sim {\cal O}({\rm several~fm})\lesssim R,R_W$, 
which confirms the importance of the screening effect.
We use the relaxation method to solve 
Eqs.~(\ref{e:poisson}--\ref{e:chemeq}) consistently.
The details of the numerical procedure to calculate the EOS and determine the
geometrical structure of the MP 
are similar to that in Refs.~\cite{mixtat,mixmaru}.

\subsection{Charge screening effect and the Maxwell construction}

\begin{figure}
\includegraphics[width=0.48\textwidth]{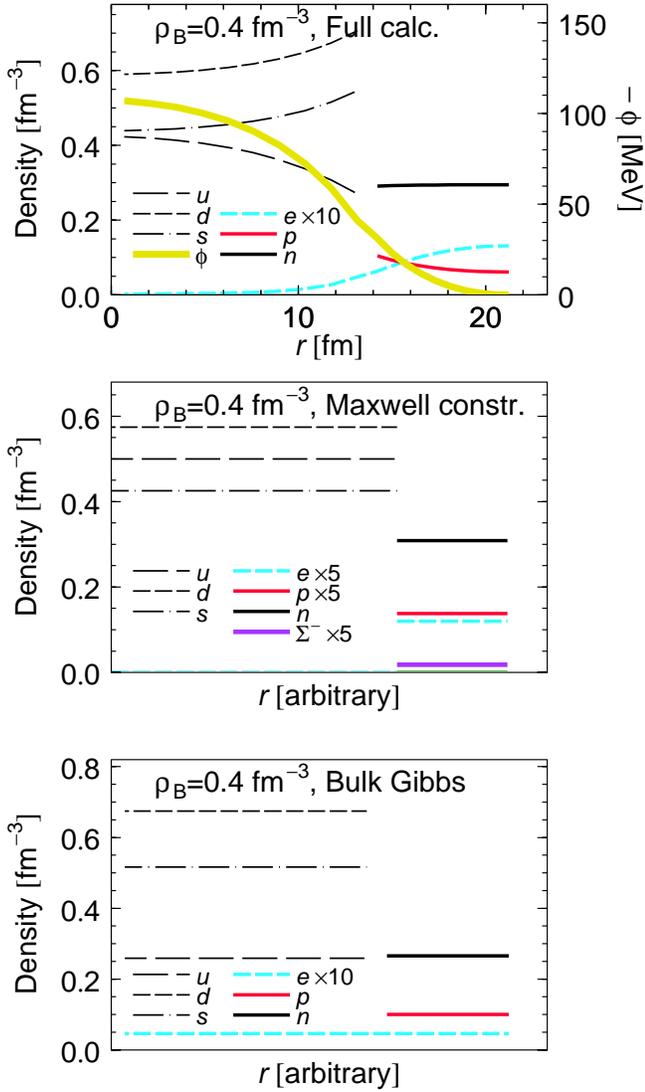}
\caption{
(Color online)
Upper panel:
Density profiles 
and Coulomb potential $\phi$ 
within a 3D (quark droplet) Wigner-Seitz cell
of the MP at $\rho_B=0.4$ fm$^{-3}$.
The cell radius and the droplet radius are $R_W=21.6$ fm
and $R=13.8$ fm, respectively.
Middle panel:
Density profiles in the MC case, where only the volume fraction is calculated,
while the absolute values of $R$ and $R_W$ are irrelevant.
We use the same $R_W$ as in the upper panel, while the volume fraction
of the quark phase is different, $\fv=0.375$.
See Fig.~\ref{figRad} for the volume fraction.
The results are for the $\bc$,
$\alpha_s=0$, $\sigma=40\;\rm MeV\!/fm^2$
case.
Lower panel:
Density profiles in the bulk Gibbs calculation.
}
\label{figProf}
\end{figure}

Figure~\ref{figProf} shows an example of the density profile in a 3D cell 
for $\rho_B=0.4\;\rm fm^{-3}$ and the $\bc$ case.
One observes a non-uniform density distribution of each particle species
together with the finite Coulomb potential, while bulk calculations use
a uniform density without the Coulomb interaction \cite{mit,glen,mix}.
This is due to the charge screening effect: the charged particle
distribution is rearranged to give 
smaller Coulomb interaction energy. 
Different from the MC case (middle panel), 
where local charge neutrality is implicitly assumed, 
the quark phase is negatively charged, so that 
$d$ and $s$ quarks are repelled to the phase boundary, 
while $u$ quarks gather at the center.
Thus local charge neutrality is obviously violated at any point 
inside the cell, even at the center of the droplet or
at the boundary of the cell.
The hadron phase is positively charged: 
protons are attracted by the negatively charged quark lump, 
while electrons are repelled. 
We shall see that such rearrangement gives rise to a remarkable effect
on the energy of the MP.

\begin{figure}
\includegraphics[width=0.48\textwidth]{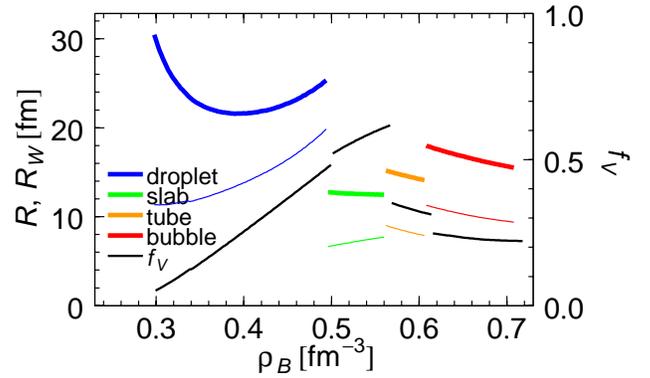}
\caption{
(Color online)
Wigner-Seitz cell radius $R_W$ (upper colored segments), 
droplet radius $R$ (lower colored segments), 
and volume fraction $\fv$ (black segments)
as a function of baryon density.
The results are for the $\bc$,
$\alpha_s=0$, $\sigma=40\;\rm MeV\!/fm^2$
case.}
\label{figRad}
\end{figure}

The density dependence of the optimal structures and their characteristic
dimensions $R$, $R_W$, and the volume fraction $\fv$ 
are shown in Fig.~\ref{figRad}.
One observes a transition from droplet to slab to tube to bubble
with increasing density.
With $\bc$ and $\beff$, 
the MP occurs in the interval
$\rho_B=0.298$ -- $0.708\;\rm fm^{-3}$ and
$\rho_B=0.236$ -- $0.670\;\rm fm^{-3}$, respectively,
i.e., it appears at less than twice normal nuclear density
and extends up to much larger density, 
relevant for NS physics.
The transitions between the different geometric structures are 
by construction discontinuous 
and a more sophisticated approach would be required for a more 
realistic description of this feature.

In the lower panel of Fig.~\ref{figProf}, shown is the case of bulk Gibbs calculation.
The local charge neutrality is not realized as in the full calculation.
However, by ignoring the Coulomb interaction, the density distribution
in each phase is uniform.

\begin{figure}
\includegraphics[width=0.48\textwidth]{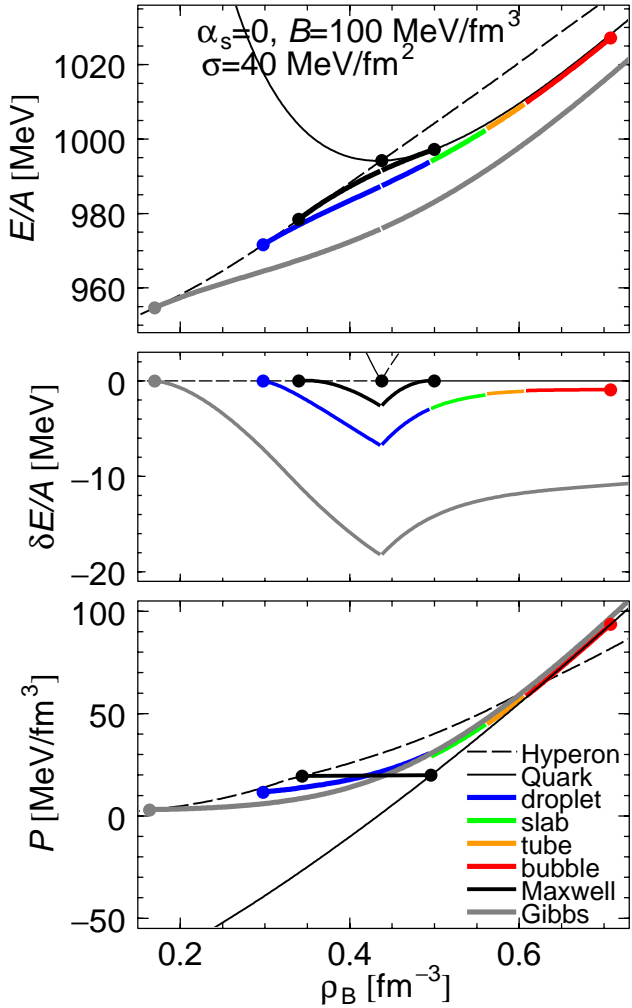}
\caption{
(Color online)
EOS of the MP (thick curves)
in comparison with pure hadron and quark phases (thin curves).
The upper panel shows the energy per baryon $E/A$, 
the middle panel the energy difference between mixed and
hadron ($\rho_B<0.44\;\rm fm^{-3}$)
or quark ($\rho_B>0.44\;\rm fm^{-3}$) phases,
and the lower panel the pressure.
Different colored segments of the MP are chosen by minimizing the energy.
The EOS within the MC 
(between $\rho_H=0.34\;\rm fm^{-3}$ and $\rho_Q=0.50\;\rm fm^{-3}$) 
is also shown for comparison
(thick black curves).}
\label{figEOS}
\end{figure}

\begin{figure}
\includegraphics[width=0.48\textwidth]{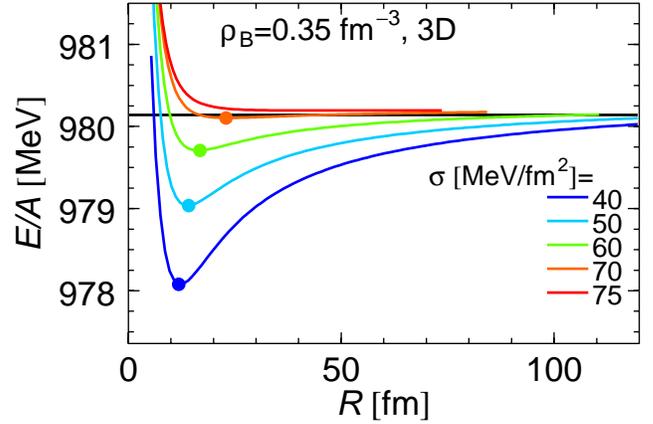}
\caption{
(Color online)
Droplet radius ($R$) dependence of the energy per baryon
for fixed baryon density $\rho_B=0.35$ fm$^{-3}$
and different surface tensions.
For any value of $R$, the density profiles and the cell size $R_W$
are optimized.
Dots on the curves show the local energy minima.
The black line shows the energy of the MC case.
}
\label{figRdep}
\end{figure}

Figure~\ref{figEOS} (upper panel) compares for $\bc$ 
the resulting energy per baryon of 
the hadron-quark MP with that of the pure hadron and quark phases
over the relevant range of baryon density. 
The thick black curve indicates the case of the MC,
while the colored line indicates the MP 
in its various geometric realizations by the full calculation.
While the structure and the composition of the MP by the full calculation 
are very different from those of the MC case, as shown before,
the energy of the MP is only slightly lower than that of the MC, 
and the resulting EOS is similar to the MC one.

The pressure is presented in the lower panel of Fig.~\ref{figEOS}. 
In the MC case, where local charge neutrality is implicitly assumed, 
the two equilibrium densities
$\rho_H\approx 0.34\;\rm fm^{-3}$ and
$\rho_Q\approx 0.50\; \rm fm^{-3}$ 
of the pure phases are connected by a straight line with equal pressure
$P_H(\rho_H)=P_Q(\rho_Q)$.
Compared with the MC,
the pressure of the MP smoothly interpolates the pressures of
the hadron and quark phases and is no more constant. 
We shall see that the
difference between the MC and our 
full calculation
causes no serious difference regarding the bulk properties of
compact stars like the mass-radius relation. 
However, the internal properties are very different between the two cases, 
which may affect the microscopic elementary processes 
in the MP \cite{red}.

If one uses a smaller surface tension parameter $\sigma$, 
the energy gets lower and the density range of the MP gets wider.
The limit of $\sigma=0$ leads to a bulk application of 
the Gibbs conditions without the Coulomb and surface effects, i.e.,
the so-called Glendenning construction \cite{gle}.
On the other hand, using a larger value of $\sigma$, 
the geometrical structures increase in size and
the EOS gets closer to that of the MC case. 
Above a critical value of $\sc$ 
the structure of the MP becomes mechanically unstable \cite{vos}: 
for a fixed volume fraction $(R/R_W)^3$
the optimal values of $R$ and $R_W$ go to infinity
and local charge neutrality is recovered in the MP, 
where the energy density equals that of the MC 
(see Fig.~\ref{figRdep}). 

This mechanical instability is 
a charge screening effect:
The optimal values of $R$ and $R_W$ are basically determined by the
balance between the Coulomb energy ($\sim R^2$ in the 3D case) 
and the surface energy ($\sim R^{-1}$).
However, if the charge screening is taken into account, the contribution of 
the screened Coulomb potential $\phi$ is strongly reduced 
when $R,R_W \rightarrow \infty$. 
A careful analysis by Voskresensky et al.~showed that the Coulomb 
energy changes its sign and behaves like
$R^{-1}$ as $R \rightarrow\infty$ 
due to the charge screening 
\cite{vos}.
Thus the surface and the Coulomb energy give a local energy minimum below $\sc$,
which disappears when the surface energy
becomes greater than the Coulomb energy above $\sc$.
This is in contrast to the work of Heiselberg et al.~\cite{mix},
neglecting the charge screening effect,
where there is always a local energy minimum at finite $R$.
The importance of the charge screening effect has also been
discussed in the context of the stability of strangelets \cite{hei}.

If we assume uniform density profiles
with a given volume fraction $\fv = (R/R_W)^d$ 
and the difference $\delta\qc = e(\qc^{(H)}-\qc^{(Q)})$
of the charge density between the two phases,
it is easy to see how the optimal size of the lump is determined
from the competition between the Coulomb and surface energy 
contributions \cite{gle,rav}: 
The Coulomb interaction energy is in this case
\be
 {E_C\over V_W} = 2\pi^2 (\delta\qc)^2 R^2
 {\fv\over d+2}
 \left[ \fv + {2 - d \fv^{1-2/d} \over d-2} \right] \:,
\ee
while the surface energy is simply 
$E_S/V_W = d \sigma \fv/R$. 
Hence there is always one energy minimum at finite $R$ 
for a given $\fv$ as a consequence of the balance between 
the Coulomb energy and the surface energy \cite{mix}.

However, when the Coulomb screening is present, 
the charge density is no more uniform, see Fig.~\ref{figProf},
but is rearranged to attain 
smaller Coulomb interaction energy. 
Consequently the Coulomb potential becomes short range 
due to the Debye screening of the charged particles. 
The contribution of this many-body effect to the energy is then twofold: 
one is the direct contribution of the Coulomb interaction energy 
and the other indirectly via the rearrangement effect of the
charge density, which is called the correlation energy in Ref.~\cite{vos}. 
Considering $\rho_a(\rv)$ as an implicit function of $\phi(\rv)$,
the kinetic and strong interaction energies can be expanded with respect
to the particle densities around their uniform values,
\be
 E_{\rm corr} = E_{\rm corr}^0
 +\!\!\!\! \sum_{i={n,p,\la,\sgm \atop u,d,s}} \int_{V_W}\!\!d^3\rv
 \mu_i^0 (\rv)  \left[ \rho_i(\rv) - \rho_i^0 \right]
 \; +\, \cdots \:,
\ee
where $E_{\rm corr}^0 = E_H^0 + E_Q^0$, 
$\mu_i^0$, and $\rho_i^0$
are the quantities for the system with 
uniform densities in the absence of the Coulomb screening. 
Thus the correlation energy gives rise to a new $R$-dependence 
in the energy density.
The analysis of Voskresensky et al.~showed that the contributions 
from the screened Coulomb potential and the correlation energy 
exhibit a $R^{-1}$ dependence for large $R$, 
and that both have different signs \cite{vos}. 
When the surface energy is added, we 
can easily see that the energy density has a local minimum at finite $R$
as long as $\sigma$ is not too large. 
Actually this local minimum disappears above a critical value $\sc$, 
which implies a mechanical instability of the structure in the MP.

\subsection{Hyperons in the mixed phase}

\begin{figure}
\includegraphics[width=0.48\textwidth]{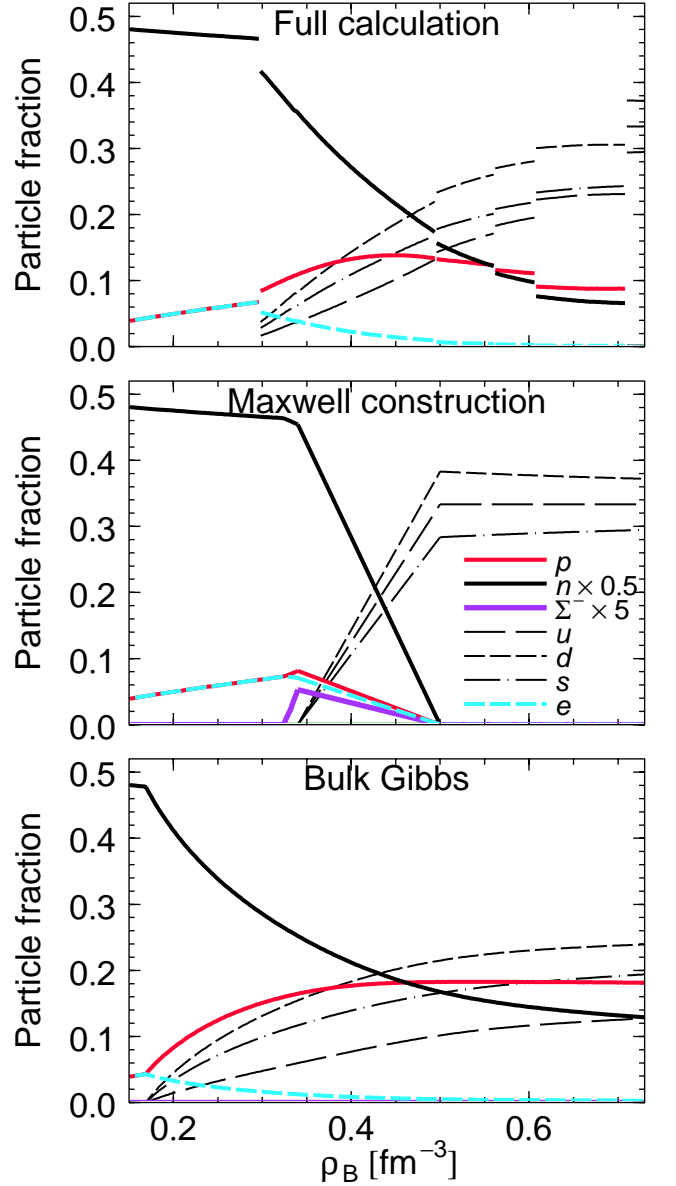}
\caption{
(Color online)
Particle fractions 
of quark and hadron species in the MP by the full calculation (upper panel),
the MC (middle panel), and the Bulk Gibbs construction (lower panel).
In the MC the phase transition occurs between the pure phases with
$\rho_H=0.34$ fm$^{-3}$ and $\rho_Q=0.50\;\rm fm^{-3}$.}
\label{figRatio}
\end{figure}

One notes in Fig.~\ref{figProf}
that no hyperons appear in the MP
although the mean baryon density $\rho_B=0.4\;\rm fm^{-3}$ 
is higher than the threshold density for hyperons 
in pure nucleon matter, $\rho_B\approx0.34\;\rm fm^{-3}$ 
(see the black dot in Fig.~\ref{figParam}).
In fact in the MC a small fraction of $\sgm$ hyperons is present,
as displayed in the lower panel of Fig.~\ref{figProf}.

This is confirmed in Fig.~\ref{figRatio}, where we compare the 
particle fractions as a function of baryon density in the full calculation (upper panel),
the MC (middle panel), and the Bulk Gibbs construction (lower panel).
One can see that the compositions are very different in the three cases,
the MP by the full calculation lying in between the extreme cases of Bulk Gibbs and MC.
In particular, a relevant hyperon ($\sgm$) fraction is only present in the 
hadronic component of the MC.
In this case the phase transition occurs at constant pressure between
the pure hadron and quark phases 
with the equilibrium densities $\rho_H=0.34$ fm$^{-3}$
and $\rho_Q=0.50\;\rm fm^{-3}$,
and for a given $\rho_B$ the volume fraction of the quark phase is
$0 \leq \fv = (\rho_B-\rho_H)/(\rho_Q-\rho_H) \leq 1$.
Accordingly the hyperon number fraction is always finite in the MC,
but in the full calculation and the Bulk Gibbs case hyperons are completely suppressed.

\begin{figure}[t]
\includegraphics[width=0.48\textwidth]{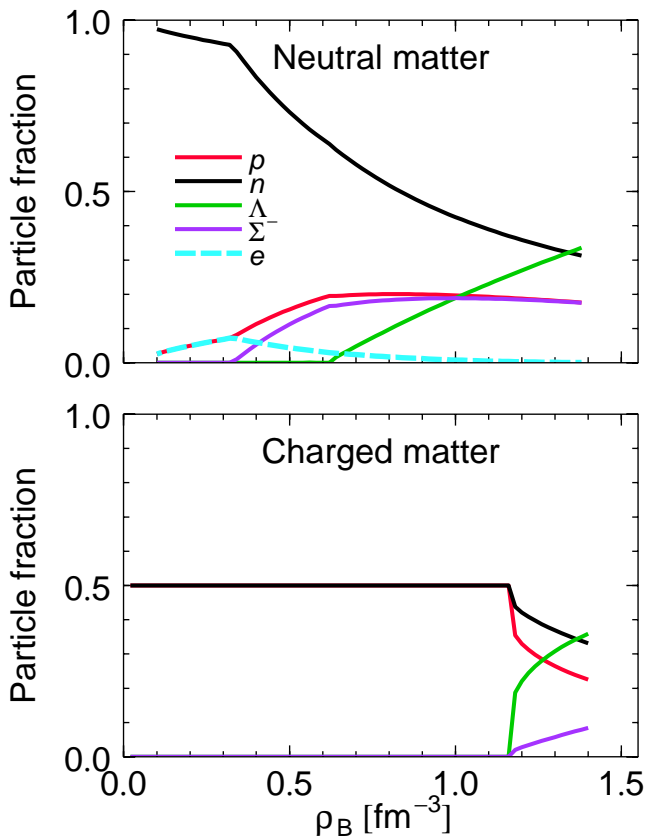}
\caption{
(Color online)
Upper panel:
Particle fractions of neutral matter with electrons (corresponding to neutron star matter).
Lower panel:
The same quantity for charged matter without electrons,
the low-density part of which corresponds to symmetric nuclear matter.
Both cases require chemical-equilibrium condition.
}
\label{figRatioUnif}
\end{figure}

This hyperon suppression is due to the absence of the
charge-neutrality condition in each phase.
In a charge-neutral hadron phase,
hyperons ($\sgm$) appear at $\rho_B=0.34\;\rm fm^{-3}$.
This is to reduce the electron Fermi energy by replacing the 
negative charge of electrons with that of $\sgm$ particles.
In the MP, on the other hand, the hadron phase can be
positively charged due to the presence of the negatively charged quark phase.
This causes the reduction of electron density and chemical potential
in the hadron phase and, consequently, $\sgm$ is suppressed.
In other words, with the charge-neutrality condition hyperons appear at 
low density to reduce the energies of electrons and neutrons in spite of 
large hyperon masses (see Fig.~\ref{figRatioUnif}).
Without charge-neutrality condition, hyperons appear at higher density 
due to their large masses.

Thus we conclude that
due to the relatively small magnitudes of the surface and Coulomb energies,
the EOS of the MP is similar to the MC one,
but the internal structure of the MP is very different.
In particular the role of hyperons is strongly reduced when we consider the
deconfinement transition in hyperonic matter.
Above a critical value of the surface tension parameter, however, 
the MC is effectively recovered as the physical phase transition.
These results should be important for physical processes
like neutrino propagation and baryonic superfluidity,
besides the maximum mass problem, which will be discussed now.

\section{Hybrid star structure}
\label{s:ns}

Knowing the EOS comprising hadronic, mixed, and quark phase
in the form $P(\eps)$,
the equilibrium configurations of static NS are obtained
in the standard way
by solving the Tolman-Oppenheimer-Volkoff (TOV) equations \cite{ns} for 
the pressure $P(r)$ and the enclosed mass $m(r)$,
\begin{eqnarray}
  {dP\over dr} &=& -{ G m \epsilon \over r^2 } \,
  {  \left( 1 + {P / \epsilon} \right) 
  \left( 1 + {4\pi r^3 P / m} \right) 
  \over
  1 - {2G m/ r} } \:,\qquad
\\
  {dm \over dr} &=& 4 \pi r^2 \epsilon \:,
\end{eqnarray}
being $G$ the gravitational constant. 
Starting with a central mass density $\epsilon(r=0) \equiv \epsilon_c$,  
one integrates out until the surface density equals the one of iron.
This gives the stellar radius $R$ and its gravitational mass $M=m(R)$.
For the description of the NS crust, we have joined 
the hadronic EOS with the ones by 
Negele and Vautherin \cite{nv} in the medium-density regime, and the ones   
by Feynman-Metropolis-Teller \cite{fey} 
and Baym-Pethick-Sutherland \cite{baym} for the outer crust.

\label{s:res}

\begin{figure}[t]
\includegraphics[width=0.48\textwidth]{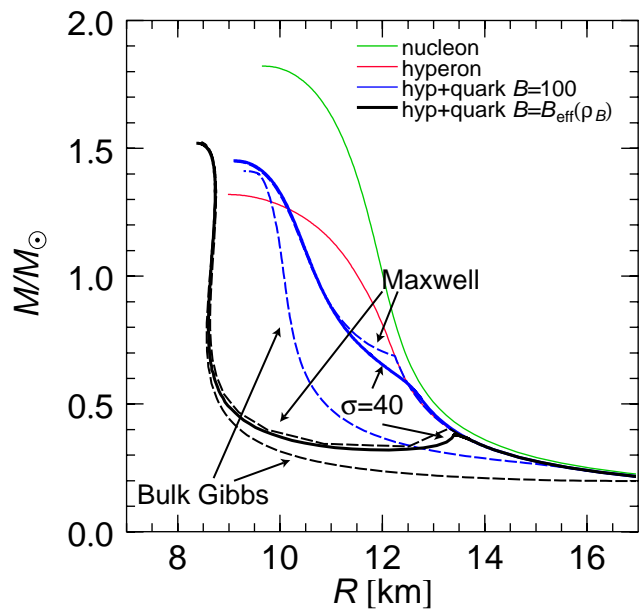}
\caption{
(Color online)
Neutron star mass-radius relations for different EOS 
and three different hadron-quark phase transition constructions.
For the hybrid stars (blue and black curves),
the dashed lines indicate the Maxwell (upper curves)
or bulk Gibbs (lower curves) constructions
and the solid lines 
the mixed phase by the full calculation.
}
\label{f:mr}
\end{figure}

\begin{figure*}[t]
\includegraphics[width=0.98\textwidth]{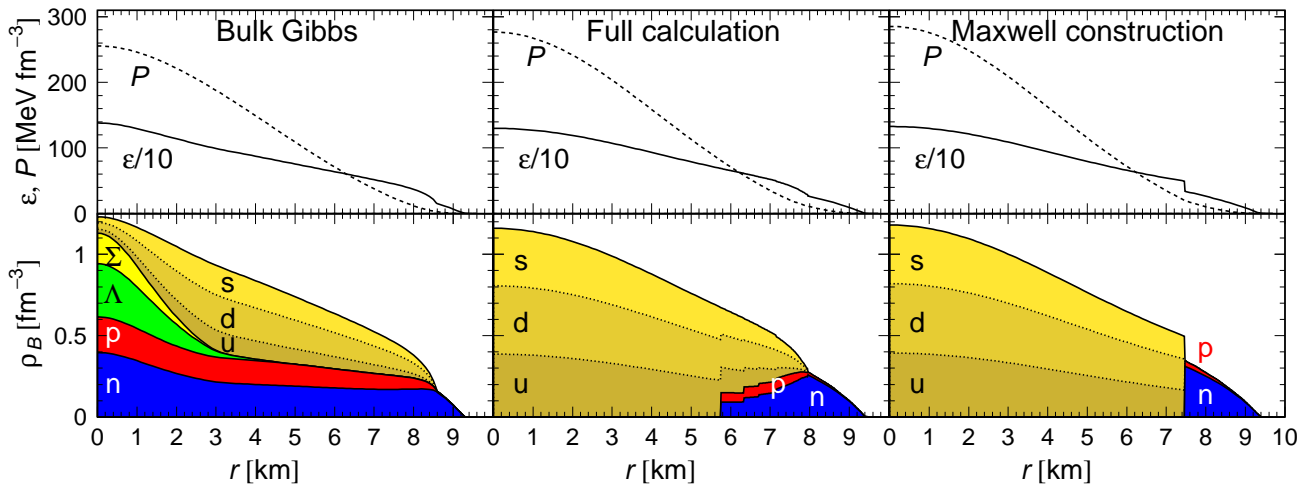}
\caption{
(Color online)
Internal structure of a $1.4\,\ms$ neutron star
obtained with three different phase transition constructions.
The upper panels show total energy density and pressure
and the lower panels the overall particle fractions as functions
of the radial coordinate of the star,
using the bulk Gibbs (left panel), 
the mixed phase by the full calculation (central panel),
and the Maxwell construction (right panel).
}
\label{f:xr}
\end{figure*}

Fig.~\ref{f:mr} compares the mass-radius relations obtained with the 
different models.
The purely nucleonic EOS (green curve) yields a maximum NS mass
of about $1.82\,\ms$, which is reduced to $1.32\,\ms$ when allowing
for the presence of hyperons (red curve).
This feature has been shown to be fairly independent of the nucleonic 
and hyperonic EOS that are used \cite{hypns2}.
The canonical NS with mass of about $1.4\,\ms$ 
can therefore not be purely hadronic stars in our approach.
In fact, the inclusion of quark matter augments the maximum mass 
of hybrid stars to about $1.5\,\ms$.

More precisely, we compare in the figure results obtained with the two quark EOS
$\bc$ (blue curves) and $\beff$ (black curves), and involving the 
different phase transition constructions Bulk, Mixed, and Maxwell.
In general, the Maxwell construction leads to a kink in the $M(R)$ relation,
because the transition from a hadronic to a hybrid star occurs suddenly,
involving a discontinuous increase of the central density when the quark
phase starts in the core of the star.
The Bulk construction yields smooth mass-radius relations 
involving a continuous transition from a hadronic to a hybrid star 
beginning at rather low central density corresponding to very low NS mass.

The MP construction by the full calculation lies between the two extreme cases,
and with our choice of $\sigma=40\;\rm MeV\!/fm^2$
it is rather close to the Maxwell construction,
smoothing out the kink of the hadron-hybrid star transition.
This transition occurs generally at a fairly low NS mass,
even below the natural minimum mass limit due to the 
formation via a protoneutron star \cite{proto}
and is thus an unobservable feature.

On the contrary, the maximum mass is hardly affected by the
type of phase transition:
For the $\beff$ model the maximum mass is $1.52\,\ms$, 
practically independent of the kind of phase transition, 
whereas for the $\bc$ model there is a slight variation of 
$M=1.45,1.45,1.50\,\ms$ for the Maxwell, mixed, and bulk construction,
respectively.

Whereas the maximum masses are practically independent of the phase 
transition construction, there are evidently large differences for the internal
composition of the star.
This is illustrated in Fig.~\ref{f:xr}, which shows the 
total energy density, pressure, and
particle fractions as a function of the radial coordinate for a
$1.4\,\ms$ NS.
One observes with the bulk Gibbs construction (left panel)
a coexistence of hadrons and quarks throughout the whole
interior of the star, 
whereas with the MC (right panel)
an abrupt transition involving a discontinuous 
jump of energy and baryon density occurs at a distance 
$r\approx7.4\;\rm km$ from the center of the star.
The small contamination with $\sgm$ hyperons in the hadronic phase
is not visible on the scale chosen.
The MP with the full calculation (central panel) lies between the two extreme cases, 
hadrons and quarks coexisting in the range $r\approx5.7$ -- $8.0\;\rm km$.
In both latter cases the pure quark matter core has a higher central pressure
and baryon number and energy densities 
than the mixed core of the first case.

\section{Summary and concluding remarks}
\label{s:end}

In this article we have studied the properties of the mixed phase 
in the quark deconfinement transition in hyperonic matter, and their
influence on compact star structure.
The hyperonic EOS given by the BHF approach with realistic hadronic
interactions is so soft that the transition density becomes 
very low if one uses the MIT bag model for the quark EOS. 

The hyperon-quark mixed phase was consistently treated with the
basic thermodynamical requirement due to the Gibbs conditions. 
We have seen that the resultant EOS
is little different from the one given by the Maxwell construction. 
This is because the finite-size effects, the surface tension, 
and the Coulomb interaction tend to diminish the available density region 
through the mechanical instability,
as has also been suggested in previous articles \cite{vos,mixtat}. 

For the bulk properties of compact stars, such as mass or radius, 
our EOS gives similar results as those given by the Maxwell construction. 
The maximum mass of a hybrid star is around $1.5\,M_\odot$, larger than
that of the purely hyperonic star, $\approx1.3\,M_\odot$. 
Hence we may conclude that a hybrid star is still consistent with the 
canonical NS mass of $1.4\,M_\odot$, 
while the masses of purely hyperonic stars lie below it.

On the other hand, the internal structure of the mixed phase is very
different; e.g., the charge density as well as the baryon number density
are nonuniform in the mixed phase. 
We have also seen that the hyperon number fraction
is 
suppressed in the mixed phase due to the relaxation
of the charge-neutrality condition,
while it is always finite in the Maxwell construction.
This has important consequences for the elementary processes inside compact
stars. 
For example, coherent scattering of neutrinos off lumps in the
mixed phase may enhance the neutrino opacity \cite{red}. 
Also, the absence of hyperons prevents a fast cooling mechanism by way of 
the hyperon Urca processes \cite{pet}. 
These results directly modify the thermal evolution of compact stars.

Although we have considered the phase transition at zero temperature, 
our study can be easily extended to finite temperature, which is 
relevant to protoneutron stars and supernovae.

In this article we have not included hyperon-hyperon interactions
and three-body forces among hyperons and nucleons, since there are still
many theoretical and experimental ambiguities.  
However, some works have suggested their relevance for the maximum 
mass problem \cite{nis}: 
if the hadronic EOS is sufficiently stiffened by repulsive interactions, 
the maximum mass problem may be resolved. 
Even in this case, however, the quark deconfinement
transition may occur and the properties of the mixed phase deserve
further investigation.

Finally, we have considered here a very simple quark matter model 
based on the MIT bag model, but there are many works about the properties of
high-density QCD.
Since color superconductivity \cite{alford} or magnetism \cite{tat} 
in quark matter are closely related to the thermal and magnetic evolutions
of compact stars, it should be interesting to take into account these
effects in the quark EOS for a more realistic description of the mixed phase. 
For example, one may expect 2SC in the quark phase, as inferred from
Fig.~\ref{figProf}: 
the number densities of $u$ and $s$ quarks become similar in the
mixed phase, while the quark densities are well different in the uniform
quark matter \cite{alford,alf,alf2}. 
It would be an interesting possibility and worth studying in detail, 
but lies outside the scope of the present paper.

\section{Acknowledgments}

We would like to acknowledge valuable discussions with 
T.~Takatsuka, R.~Tamagaki, S.~Nishizaki, and T.~Muto.
T.~M.~is grateful to the Nuclear Theory Group and the RIKEN BNL Research Center
at Brookhaven National Laboratory
for their warm hospitality and fruitful discussions.
This work is partially supported by the
Grant-in-Aid for the 21st Century COE
``Center for the Diversity and Universality in Physics''
and the Grant-in-Aid for Scientific Research Fund (C)
of the Ministry of Education, Culture, Sports, Science, and Technology of Japan
(13640282, 16540246).


\end{document}